\documentclass[a4paper,twoside]{article}

\usepackage{epsfig}
\usepackage{subcaption}
\usepackage{calc}
\usepackage{amssymb}
\usepackage{amstext}
\usepackage{amsmath}
\usepackage{amsthm}
\usepackage{multicol}
\usepackage{pslatex}
\usepackage{apalike}
\usepackage{amssymb}
\usepackage{multirow}
\usepackage{float}
\usepackage{stfloats}
\usepackage{textcomp}
\usepackage{url}
\usepackage{SCITEPRESS}     

\begin{document}

\title{Detecting message modification attacks on the CAN bus with Temporal Convolutional Networks}

\author{
 \authorname{
     Irina Chiscop\sup{1}\orcidAuthor{0000-0002-1249-8518}, 
     András Gazdag\sup{2}\orcidAuthor{0000-0002-4481-3308},
     Joost Bosman\sup{1}\orcidAuthor{0000-0001-6325-1462}, and
     Gergely Biczók\sup{2}\orcidAuthor{0000-0002-3891-3855}
 }
 \affiliation{\sup{1}Cyber Security \& Robustness Department, TNO, The Hague, The Netherlands}
    \affiliation{\sup{2}CrySyS Lab, Dept. of Networked Systems and Services, Budapest University of Technology and Economics, Budapest, Hungary}    
 \email{\{irina.chiscop\}@tno.nl, \{agazdag,biczok\}@crysys.hu}
}


\abstract{Multiple attacks have shown that in-vehicle networks have vulnerabilities which can be exploited. Securing the Controller Area Network (CAN) for modern vehicles has become a necessary task for car manufacturers.
Some attacks inject potentially large amount of fake messages into the CAN network; however, such attacks are relatively easy to detect. In more sophisticated attacks, the original messages are modified, making the detection a more complex problem.
In this paper, we present a novel machine learning based intrusion detection method for CAN networks.  We focus on detecting message modification attacks, which do not change the timing patterns of communications. Our proposed temporal convolutional network-based solution can learn the normal behavior of CAN signals and differentiate them from malicious ones. The method is evaluated on multiple CAN-bus message IDs from two public datasets including different types of attacks. Performance results show that our lightweight approach compares favorably to the state-of-the-art unsupervised learning approach, achieving similar or better accuracy for a wide range of scenarios with a significantly lower false positive rate.}

\onecolumn \maketitle \normalsize \setcounter{footnote}{0} \vfill

\section{\uppercase{Introduction}}
\label{sec:introduction}
A modern automobile contains more than 100 electronic control units (ECU) to control the vehicular  subsystems and help the driver with various sophisticated services. These ECUs are spread over the entire vehicle and connected mostly via Controller Area Networks (CANs). CAN was initially designed to be an isolated system decades ago, however, in the age of round-the-clock connected networked objects, this property no longer holds. Realizing the aforementioned connectivity, Bluetooth, Wi-Fi, or cellular connections are all potential intrusion points for an attacker. If a malicious actor compromises a component\footnote{\url{www.wired.com/2015/07/hackers-remotely-kill-jeep-highway}}\textsuperscript{,}\footnote{\url{www.wired.com/story/tesla-model-x-hack-bluetooth}} that implements one of these connections, it becomes possible to manipulate the CAN network. Dashboard information is displayed and actuators in Advanced Driver-Assistance Systems (ADAS) are controlled based on sensor readings transmitted over the CAN bus; interfering with these messages may result in significant financial loss and danger to human life.

Two approaches have so far been seen among the attacks. 
Existing attacks are of two distinct types: (i) the attacker either injects additional CAN messages into the network or (ii) she modifies otherwise valid messages sent by legitimate ECUs. We note that the latter attack is very difficult to implement by altering electrical signals of the CAN on-the-fly, yet it can be realized by compromising an ECU and sending out messages with modified content, or by compromising a gateway between two CAN networks and modifying messages passing through it  \cite{Gazdag2020}. Message injection attacks can be detected easily because the original messages are also present next to malicious ones, which changes the temporal patterns of traffic. On the other hand, detecting modification attacks poses a tougher challenge: it requires an in-depth analysis of the actual payload as the rest of the traffic properties remain intact in this scenario.

\noindent\textbf{Related work. }In recent years, a considerable amount of literature has been published on CAN bus intrusion detection. These works can be split into three categories: frequency-, statistics-, or machine learning based methods. Most of these approaches are particularly useful for detecting cyber-attacks in which additional messages are being injected into the CAN bus.
The simplest of the three, frequency-based models focus on testing inter-arrival times of CAN messages against a predefined normal baseline \cite{Taylor2015,Song2016,Moore2017,Gazdag2018}. As the name suggests, statistics-based detection approaches exploit the statistical properties of CAN bus traffic such as entropy \cite{Muter2011}, Z-score \cite{Tomlinson2018} or Mahalanobis distance \cite{Xiuliang2019}.
Machine learning based methods imply the usage of artificial neural networks, clustering and supervised models for classification and regression. In the specific field of CAN bus intrusion detection, popular machine learning approaches include autoencoders \cite{Lokman2019,Lin2020,Novikova2020}, recurrent neural networks (RNN) such as Long Short-Term Memory (LSTM) networks~\cite{Taylor2016,Negi2019,Khan2020,Hanselman2020,Hossain2020}, Gated Recurrent Unit (GRU)-based networks~\cite{Kumar2020}, replicator neural networks \cite{Weber2018}, and deep convolutional networks~\cite{Song2020}. The scrutinized literature shows that recurrent architectures are often the preferred choice for modeling the time series of CAN bus signals, whilst convolutional networks are used when data is transformed to a two-dimensional grid dataframe to resemble an image format~\cite{Song2020}. In particular, only one approach was found to combine these two techniques in the form of a convolutional LSTM \cite{Tariq2020} which is trained on labeled data in a supervised fashion.

\noindent\textbf{Temporal Convolutional Networks.} To the best of our knowledge, no existing solution employs (causal) convolutions to model the time series representation of CAN signals; we argue that such an approach makes perfect sense given the successful application of convolutional networks to sequence modeling tasks. Specifically, a Temporal Convolutional Network (TCN) is a type of convolutional network whose architecture consists of causal (and dilated) convolutions~\cite{TCN2018}. It has been shown that this new type of network outperforms recurrent architectures, such as LSTM and GRU, on a multitude of sequence modeling tasks including the adding problem and image classification on sequential MNIST and P-MNIST~\cite{TCN2018}. In fact, TCNs have also been successfully applied to anomaly detection in general time series data \cite{He2019}.

\noindent\textbf{Our contribution.} In this paper, we propose a TCN-based approach for detecting modified CAN bus messages; our focus is solely on message modification attacks with no message injection.  We construct and train the TCN in an unsupervised fashion, since, in practice, labelling CAN bus messages is a very difficult task. In the training process, the TCN will learn to accurately reconstruct the signals of individual CAN bus messages through its causal convolution layers, which allows for information retention from past data samples. Finally, the classification of new data samples will resume to setting an appropriate threshold on their reconstruction loss value. The core idea here is that signals whose data have been altered will be poorly reconstructed by the model, and thus be easy to recognize. Note, that it is not a prerequisite for us to know CAN bus signal semantics which varies for vehicle make and model, and is usually kept confidential~\cite{Lestyan2019,Remeli2019}.
The contribution of this paper is three-fold:
\begin{enumerate} 
    \item{We first introduce a new CAN bus dataset containing both benign data and synthetic attacks.} 
    \item{We then propose a TCN architecture to learn and reconstruct the normal behaviour of CAN bus signals, and use this information to pinpoint anomalies that do not conform to the reconstruction given by model.} 
    \item {We compare the detection performance of our approach to a state-of-the-art GRU-autoencoder~\cite{Kumar2020} (shown to outperform other existing solutions) through numerical experiments on both our own dataset and the \emph{de facto} standard SynCAN dataset \cite{Hanselman2020}. Results show that our simple TCN-based approach compares favorably to the state-of-the-art, i.e., it achieves similar or better accuracy with a significantly lower false positive rate.}
\end{enumerate}
\noindent\textbf{Paper structure. }The rest of the paper is structured as follows. Section \ref{sec:method} presents our proposed TCN architecture in detail.  Section \ref{sec:experiment} describes the design of our experiments including choosing the baseline, introducing our two datasets and the training process, and defining evaluation metrics. Section \ref{sec:results} presents the results of the comparative performance evaluation. Finally, Section \ref{sec:conclusion} concludes the paper.

\section{\uppercase{Intrusion detection model}}
\label{sec:method}

In this section we present the motivation behind choosing temporal convolutional networks as an intrusion detection mechanism for the CAN bus. We first provide some background on convolutional networks and then describe our proposed TCN architecture in detail.

\subsection{Convolutional networks}
\label{subsec:convolutions}
Convolutional neural networks are a particular kind of deep neural networks that enables the extraction of relevant spatial and temporal features from the input (e.g., an image) by learning a set of filters. These filters represent multi-dimensional arrays sliding over the input image, and are initialized randomly. During the forward pass, the dot product between the entries of each filter and the image sub-block is computed, resulting in a feature map. When another convolutional layer is added, the features learned in the first layer are combined to create new ones. To account for as many (non-linear) combinations of features as possible, it is customary to increase the filter size in the subsequent layers. The deeper the network becomes, the better it gets at extracting refined patters from the data. A more detailed description of different convolutional architectures can be found in \cite{Aloysius2017}.

\begin{figure}[ht]
\centering
\includegraphics[width=0.95\linewidth]{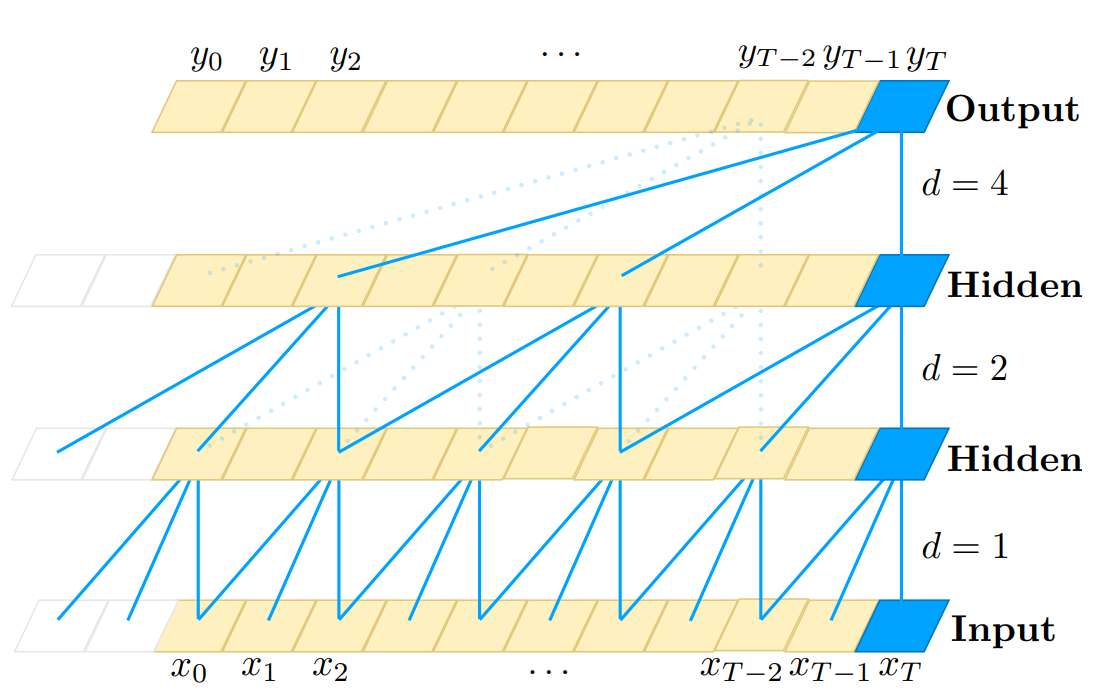}
\caption{A dilated causal convolutional neural network with two hidden layers, dilation factors $d = 1, 2, 4$ and filter size $k = 3$ \cite{TCN2018}.}
\label{fig:dilated_conv}
\end{figure}

Temporal convolutional networks (TCN) are a category of convolutional networks particularly suitable for modeling long-term dependencies in sequential data \cite{Oord2016,TCN2018}. Consider for instance the following task: based on input sequence $x_{0},x_{1},\ldots, x_{T}$, predict corresponding output $y_{0},y_{1},\ldots,y_{T}$ at each time step. There are two constraints associated with this task. First, the predicted output $y_{t}$ should only be influenced by previously observed inputs $x_{0},x_{1},\dots,x_{t}$, and, second, the size of the network output must be identical to that of the input sequence. TCNs tackle the first constraint by
sliding a filter only over the past input values. In other words, the convolution filter has positive weights only for past inputs. TCNs also employ dilated causal convolutions which, unlike regular causal convolutions, enable an exponential growth of the receptive field by skipping over the inputs while convolving. Moreover, a larger receptive field allows the neural network to infer the relationships between different observations in the input data. The second constraint is addressed by padding the input data with zeros at the borders, to control the dimension of the output. These two architectural elements can be observed in Figure~\ref{fig:dilated_conv}, depicting a dilated causal convolutional network with two hidden layers. Here, the zero-padding is represented by the white squares on the left side. The filter size of $k=3$ is indicated by the blue lines. The dilation factor $d$, applied at each layer, indicates how many input values are being skipped by the filter. Increasing the dilation factor by 2 at each subsequent layer results in a receptive field of size 15:  the value of a neuron in the output layer is influenced by fifteen neurons from the input layer. 

TCNs possess numerous advantages when compared to recurrent architectures \cite{TCN2018}. Convolutions within TCNs can be computed in parallel, thus allowing the entire data sequence to be processed. That is not possible with RNNs, where the computation of the output at a specific timestep requires the complete computation of all its predecessors. Moreover, TCNs require less memory during training than RNNs, where partial values of cell-gates need to be stored, and exhibit stable gradients, as backpropagation does not happen through multiple different time samples. In theory the receptive field of RNNs is infinite; in TCNs the field is finite, and its size depends on the number of layers (dilations) and filters used. Apparently, there exists a trade-off between how lightweight the network is, and its ability to capture long-term dependencies in the data. Both aspects are equally important to obtain a scalable and reliable CAN bus intrusion detector. In the remainder of this paper we show that a TCN model is a suitable candidate for this purpose.

\subsection{TCN architecture}
\begin{figure}[tb]
\centering
\includegraphics[width=1\linewidth]{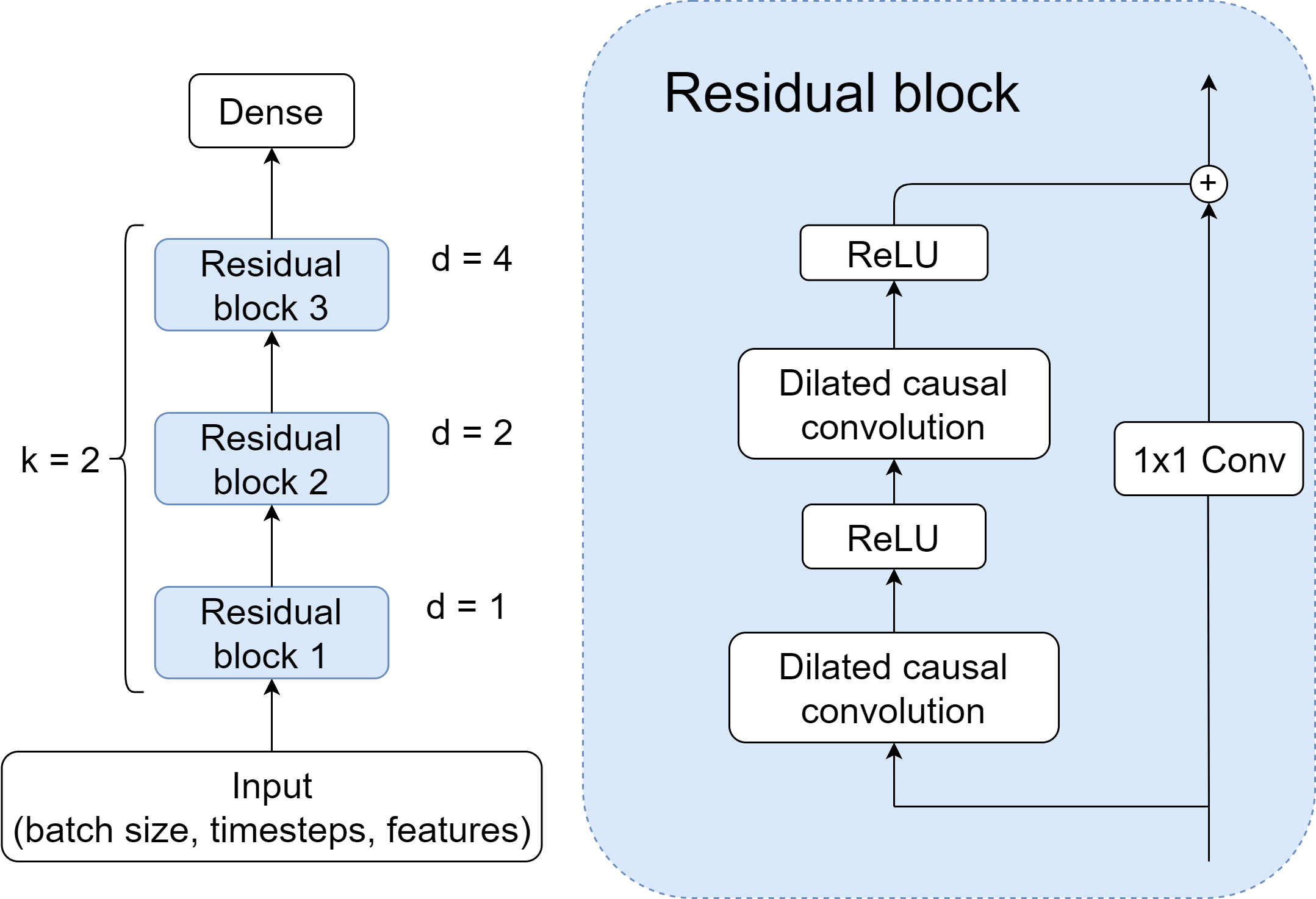}
\caption{Our TCN architecture with three residual blocks with convolutional dilations and filter size of $k=2$.}
\label{fig:TCN_architecture}
\end{figure}

The TCN to be used for CAN bus intrusion detection follows the general framework from \cite{TCN2018} and is shown in Figure~\ref{fig:TCN_architecture}.  The network consists of an input layer, three residual blocks, and an output layer. As shown in the figure, the input for the TCN must be three-dimensional. Each residual block contains two dilated causal convolution layers each having  64 filters and the same dilation factor $d$. The Rectified Linear Unit (ReLU) is used as an activation function on these layers. The filter size is kept at the same value of $k=2$ across all residual blocks. A skip connection is also enabled, which adds the output from the previous layer to the next layer. This is marked by the element-wise addition $\oplus$. Due to zero-padding, this operation may receive inputs that differ in shape. To circumvent this, a 1x1 convolution is added.

The network is kept simple deliberately: no weight normalization or dropout layers have been used. Our main objective here is to investigate whether this lightweight TCN can successfully learn to reconstruct CAN bus signals, and achieve results comparable to or better than other, more complex state-of-the-art classifiers.

\subsection{Intrusion score and output}
\label{sec:score}
We distinguish between benign and malicious messages by applying a threshold to the reconstruction loss. We therefore monitor the squared error between the signal value at a given times and its latest reconstructed value. This defines an intrusion score for each signal in a message. To compute an intrusion score per message, we calculate a set of thresholds given by the $99.9$th percentile of the validation loss for each signal in the data. A message is then labeled as malicious if one of the signal's intrusion scores exceeds the threshold set for that signal. We opted for this approach to label messages based on individual signal thresholds: in practice, depending on the complexity and correlation of the signals, some may be better reconstructed during training than others. 

\section{\uppercase{Experiment design}}
\label{sec:experiment}

In this section we describe the design of our numerical experiments, including the baseline model, datasets, training process and our choice of evaluation metrics.

\subsection{Selecting the most suitable baseline}
For evaluation purposes, we identified the best-performing CAN bus anomaly detection algorithms by scrutinizing recent literature. We used the following selection criteria:
\begin{itemize}
    \item Unsupervised learning: the algorithm requires no labeled data for training.
    \item Generalization: the algorithm is easy to generalize, and thus does not depend on data pre-processing such as identifying and pre-selecting specific CAN signals.
    \item Fully-reproducible: the algorithm needs to be accompanied by sufficient information in order to have a fully reproducible implementation. 
\end{itemize}

To the best of our knowledge, the most recent and suitable candidate is the INDRA framework \cite{Kumar2020}. It proposes a recurrent autoencoder network that is able to detect CAN messages in which signals have been tampered with. For each message ID one such recurrent autoencoder is trained such that it learns to reconstruct the signals within that particular message ID. This approach is shown to outperform other recent unsupervised methods such as Predictor LSTM \cite{Taylor2015}, Replicator Neural Network \cite{Weber2018}, and CANet \cite{Hanselman2020}, on most attack classes of the SynCAN dataset, in terms of accuracy and false positive rate. Moreover, Predictor LSTM is designed to predict the raw message data in string form, and thus does not directly fall within the scope of time-series-based intrusion detection. Note that the CANet model is also more complex since its architecture combines the LSTM models of individual messages to account for capturing the correlations between different IDs. Finally, the convolutional LSTM proposed in \cite{Tariq2020} is a promising method for predicting multi variate time series data. However, it was designed for supervised learning which requires labeled data for training and for this reason, it falls outside the scope of this paper.\\
In view of these arguments, INDRA is the most sensible baseline for comparative performance evaluation.

\subsection{Datasets}
\label{sec:datasets}
When evaluating machine learning classifiers, it is considered best practice to employ multiple datasets in order to assess the impact of the number of data samples and different features on the model's performance. Moreover, publicly available CAN bus datasets for intrusion detection are labelled differently, either per message ID or per signal. To account for both, we consider two datasets: the SynCAN dataset with message labels and the CrySyS dataset with individual signal labels.

\subsubsection{SynCAN dataset}
\label{subsec:syncan}
The SynCAN (Synthetic CAN Bus Data) dataset was introduced in \cite{Hanselman2020}, and is publicly available\footnote{\url{www.github.com/etas/SynCAN}}. The dataset contains 10 different CAN message IDs, whilst the number of signals in each ID varies between 1 and 4. Overall, the dataset covers 20 correlated signals. The training data spans approximately $16.5$ hours of traffic, while the testing data about $7.5$. Moreover, testing data includes a $0$/$1$ label per individual message, to indicate whether it is malicious or not. However, there is no indication as to which signal has been attacked within a malicious message. Since this dataset is only meant for unsupervised learning purposes, the training data does not include explicit labels. Finally, the test data is split across six different files, each corresponding to a different simulated attack:
\begin{itemize}
    \item \textit{Plateau attack:} the value of a single signal is overwritten by a constant value over a certain period of time.
    \item \textit{Continuous change attack:} the value of a signal is overwritten at a slow pace, such that it increasingly deviates from its true value.
    \item \textit{Playback attack:} the values of a signal within a time interval is overwritten with the values of the same signal from a randomly selected past interval. 
    \item \textit{Suppression attack:} signal values contained in a certain message ID simply do not appear in the CAN traffic for a period of time.
    \item \textit{Flooding attack:} messages with a certain ID are sent with a higher frequency to the CAN bus.
\end{itemize}

Detection of message injection attacks (suppression attack and flooding attack) is not a goal of this paper. Nonetheless, in Section \ref{sec:results}, we evaluated our TCN architectures performance on those as well for a better comparison with the INDRA model.

\subsubsection{CrySyS dataset}\label{subsec:crysys-data}
The CrySyS dataset was created by the CrySyS Lab in the context of the SECREDAS project~\footnote{\url{www.secredas-project.eu}}, and it is also publicly available\footnote{\url{www.crysys.hu/research/vehicle-security}}. It is significantly smaller compared to the SynCAN dataset, however, the driving environment and the behavior of the vehicle are better known. It contains 7 smaller ($<$1 minute) captures of specific driving and traffic scenarios, and a longer trace (\texttildelow 25 minutes). There are 20 different message IDs in the traces, and the number of signals varies between 1 and 6.

Additionally, to complement this dataset, we have developed a signal extractor and an attack generator script. The signal extractor is based on the work presented in \cite{Stone2018}. It calculates the statistical properties of bits in CAN data fields, and identifies and separates the signals based on the changes in these values.

The attack generator\footnote{\url{www.github.com/CrySyS/can-log-infector}} is able to modify the CAN messages. It changes some or all the values in the data field of existing messages without modifying the timing of a message. To achieve a meaningful targeted attack, the generator can be combined with the information gathered from the signal extractor to modify specific signals in the trace. The attack generator also supports multiple attack types:

\begin{itemize}
    \item \textit{Change to constant:} the original value is replaced by the given constant value.
    \item \textit{Change to random:} the original value is replaced by a new random value.
    \item \textit{Modify with delta:} the given value is added to the original data value.
    \item \textit{Modify with increment:} a per message increment is added to the original value.
    \item \textit{Modify with decrement value:} a per message decrement is subtracted from the original value.
    \item \textit{Change to increment:} the original data value is replaced by a per message incremented value.
    \item \textit{Change to decrement:} the original data value is replaced by a per message decremented value.
\end{itemize}

We modified the original CrySyS traces with the attack generator script to simulate attacks. After we identified the different signals in the traces we replaced a chosen signal with a constant value for the second half of the trace. Note that this simple change-to-constant/plateau attack was enough to demonstrate the capabilities of our approach over INDRA (see Section \ref{sec:results}). Also note that we focused on IDs with 1 to 4 signals per message, similar to SynCAN, to be able to compare the results across the two datasets.

\subsection{Training the models}
Training both the TCN and INDRA models required the normalization of signal data (values between 0 and 1), and then re-shaping the input data to three-dimensional. This was done by sliding a fixed-size window over the time series, one timestamp at a time. As in \cite{Kumar2020}, we applied a rolling window of 20 timestamps or, equivalently, of 20 messages, to the training datasets shown in Table~\ref{table:datasets}.

The rest of the training parameters were set to the same values as in \cite{Kumar2020} to ensure an accurate reproduction of the INDRA model. Concerning the optimizer and loss function, both models used the \textit{Adam} optimizer with learning rate $0.0001$ and mean square error. The models were trained for 100 epochs with a batch size of $128$ on $85\%$ of the training data, whilst the other 15\% was kept for validation. An early-stop mechanism terminated the training if the validation loss did not improve in the last 10 epochs. Note that during initial experiments, a higher number of epochs was considered, but the training stopped before the $100$th epoch in all cases. All models were implemented using the \textit{keras} and \textit{keras-tcn}\footnote{www.github.com/philipperemy/keras-tcn} libraries in Python 3.7, and trained on a GeForce GTX 960 GPU. The two models have only been trained offline, not on live CAN bus data.

\begin{table}
\centering
\caption{Overview of datasets used in the numerical experiments.}
\scalebox{0.8}{
\begin{tabular}{l|cccc}

\hline

\multicolumn{1}{c}{Dataset$\quad$} & \multicolumn{1}{c}{Message ID} & \multicolumn{1}{c}{\begin{tabular}[c]{@{}c@{}}No. of \\ signals\end{tabular}} & \multicolumn{1}{c}{
\begin{tabular}[c]{@{}c@{}}Train \\ samples\end{tabular}} & \multicolumn{1}{c}{\begin{tabular}[c]{@{}c@{}}Test \\ samples\end{tabular}} \\ 

\hline

\multirow{2}{*}{SynCAN} & 2        & 3              & 4139826       & 909869       \\
                        & 3        & 2              & 2070144       & 1884235      \\
                        & 10       & 4              & 1380087       & 610294       \\ 

\hline

\multirow{2}{*}{CrySyS} & 280        & 4              & 157472        & 3895         \\
                                 & 290      &   5            &    15748          &  389            \\ 

\hline

\end{tabular}
}
\label{table:datasets}
\end{table}

\subsection{Evaluation metrics}

To evaluate the performance of the TCN model, we use the intrusion score defined in Section \ref{sec:score}. 
The INDRA model uses the same squared error as a signal intrusion score, but applies a generic threshold set to the 99.9th percentile of the validation loss (computed across all signals). The message intrusion score is then given by the maximum signal intrusion score contained in that message, and is then compared to the threshold.
We use three standard performance metrics for the evaluation of the models: accuracy, false positive rate and precision. Accuracy measures the ratio of the predicted labels exactly matching the ground truth, and is defined as follows:
\begin{align}
    Accuracy = \frac{TP+TN}{TP+FP+TN+FN},
\end{align}
where $TN$, $TP$, $FN$, $FP$ denote the number of true, and false, positives and negatives, respectively. Accuracy gives an indication of the general classification capabilities of a certain model.

The false positive rate (FPR) measures the amount of samples wrongly classified as malicious, whilst in fact being benign. The false positive rate is extremely relevant from the practical point of view: in the CAN bus context, the messages marked as malicious may need to be further analyzed before deciding on mitigation actions. To keep operation efficient, the false positive rate needs to be minimized as much as possible. Precision, on the other hand, measures the capabilities of the model to actually detect the relevant attacks (positive samples). This is another important quantity to monitor since imbalanced datasets, with far more negatives than positives, may render accuracy a deceiving metric. In fact, CAN bus datasets are usually imbalanced, since most (simulated) attacks have a very short duration. The FPR and precision are defined as follows:
\begin{align}
    FPR = \frac{FP}{TN+FP} , \\
    Precision = \frac{TP}{TP+FP} .
\end{align}

\section{\uppercase{Results}}
\label{sec:results}
\noindent\textbf{SynCAN. }We first assessed the performance of the two models on the SynCAN dataset. The accuracy and false positive rate, calculated for the normal test set and for each attack class, are shown in Table~\ref{table:syncan_acc_fpr}. A first observation is that TCN achieves a higher accuracy than INDRA in most cases, with the exception of playback and flooding attacks on ID $10$. 
Moreover, the false positive rates are quite low for both models, which can be explained by looking at the precision values in Table~\ref{table:precision_syncan}. Overall, there are large variations in the precision values across different message IDs which may be related to how the attacks were performed (target signals chosen, attack duration, etc.) and the different signal correlations. Also, the relatively low precision values in Table~\ref{table:precision_syncan} show that the models manage to capture only a limited set of temporal characteristics of the SynCAN data. This is a direct consequence of the stopping mechanism implemented during training, and in the case of TCN, of the choices made to keep a lightweight architecture.
For playback attacks, precision is very low for both models, which leads to the similarly low false positive rates achieved in this class. This is not surprising since during a playback attack, a portion of past data is written over its current values, making the signal look normal, and thus the attack difficult to detect. TCN clearly achieves a better performance than INDRA in detecting continuous attacks. Moreover, for message IDs 2 and 3, TCN detects suppression attacks with a much larger precision compared to INDRA. This result appears to be influenced by the number of signals in the message, since precision significantly decreases as the number of signals increases. As for plateau attacks, the two methods achieve similar results. INDRA is more precise than the TCN model is on detecting flooding attacks. This is an expected result, mainly due to the TCN accurately reconstructing data from a flooding attack since the data values are not altered during such an attack. To sum up, the TCN model is capable of detecting all message modification attacks (continuous change, playback and plateau) effectively. Although detecting attacks which modify the arrival rates of CAN bus messages was not part of the original goal, TCN also proved successful at detecting suppression attacks.
\begin{table*}[t]
\centering
\caption{Results for the SynCAN dataset.}
\scalebox{0.8}{
\begin{tabular}{llcccccccccccc}
                           &                                             & \multicolumn{2}{c}{Normal}        & \multicolumn{2}{c}{Continuous}    & \multicolumn{2}{c}{Playback}      & \multicolumn{2}{c}{Flooding}      & \multicolumn{2}{c}{Suppress}      & \multicolumn{2}{c}{Plateau}       \\ \hline
\multicolumn{1}{l|}{Model} & \multicolumn{1}{l|}{Data}                   & Acc.            & FPR             & Acc.            & FPR             & Acc.            & FPR             & Acc.            & FPR             & Acc.            & FPR             & Acc.            & FPR             \\ \hline
\multicolumn{1}{l|}{TCN}   & \multicolumn{1}{l|}{\multirow{2}{*}{ID 2}}  & \textbf{0.9977} & \textbf{0.0022} & \textbf{0.8660} & \textbf{0.0018} & \textbf{0.8674} & \textbf{0.0013} & \textbf{0.7678} & \textbf{0.0026} & \textbf{0.8402} & \textbf{0.0001} & \textbf{0.8336} & \textbf{0.0066} \\
\multicolumn{1}{l|}{INDRA} & \multicolumn{1}{l|}{}                       & 0.9811          & 0.0188          & 0.8584          & 0.0121          & 0.8660          & 0.0046          & 0.7600          & 0.0157          & 0.8347          & 0.0101          & 0.8133          & 0.0495          \\ \hline
\multicolumn{1}{l|}{TCN}   & \multicolumn{1}{l|}{\multirow{2}{*}{ID 3}}  & \textbf{0.9992} & \textbf{0.0007} & \textbf{0.8664} & \textbf{0.0009} & \textbf{0.8680} & \textbf{0.0002} & \textbf{0.6422} & \textbf{0.0011} & \textbf{0.8390} & \textbf{0.0004} & \textbf{0.8394} & \textbf{0.0012} \\
\multicolumn{1}{l|}{INDRA} & \multicolumn{1}{l|}{}                       & 0.9965          & 0.0034          & 0.8653          & 0.0033          & 0.8672          & 0.0012          & 0.6420          & 0.0033          & 0.8377          & 0.0025          & 0.8386          & 0.0036          \\ \hline
\multicolumn{1}{l|}{TCN}   & \multicolumn{1}{l|}{\multirow{2}{*}{ID 10}} & \textbf{0.9977} & \textbf{0.0022} & \textbf{0.8637} & \textbf{0.0072} & 0.8577          & 0.0160          & 0.7399          & \textbf{0.0001} & \textbf{0.8446} & \textbf{0.0011} & \textbf{0.8282} & \textbf{0.0136} \\
\multicolumn{1}{l|}{INDRA} & \multicolumn{1}{l|}{}                       & 0.9858          & 0.0141          & 0.8546          & 0.0176          & \textbf{0.8638} & \textbf{0.0070} & \textbf{0.7923} & 0.0047          & 0.8370          & 0.0105          & 0.8100          & 0.0447          \\ \hline
\end{tabular}}
\label{table:syncan_acc_fpr}
\end{table*}
\begin{table*}[t]
\caption{Precision of the models on SynCAN dataset.}
\centering
\scalebox{0.8}{
\begin{tabular}{llcllcllclllllll}
\hline
\multicolumn{1}{l|}{Model} & \multicolumn{1}{l|}{Data}                   & \multicolumn{3}{c}{Continuous}      & \multicolumn{3}{c}{Playback}        & \multicolumn{3}{c}{Flooding}        & \multicolumn{3}{c}{Suppress}        & \multicolumn{2}{c}{Plateau}         \\ \hline
\multicolumn{1}{l|}{TCN}   & \multicolumn{1}{l|}{\multirow{2}{*}{ID 2}}  & \multicolumn{3}{c}{\textbf{0.4457}} & \multicolumn{3}{c}{0.2458}          & \multicolumn{3}{c}{0.0205}          & \multicolumn{3}{l}{\textbf{0.9027}} & \multicolumn{2}{l}{0.3022}          \\
\multicolumn{1}{l|}{INDRA} & \multicolumn{1}{l|}{}                       & \multicolumn{3}{c}{0.1992}          & \multicolumn{3}{c}{\textbf{0.3696}} & \multicolumn{3}{c}{\textbf{0.1577}} & \multicolumn{3}{l}{0.2812}          & \multicolumn{2}{l}{\textbf{0.3033}} \\ \hline
\multicolumn{1}{l|}{TCN}   & \multicolumn{1}{l|}{\multirow{2}{*}{ID 3}}  & \multicolumn{3}{c}{\textbf{0.5231}} & \multicolumn{3}{c}{0.0000}           & \multicolumn{3}{c}{0.1028}          & \multicolumn{3}{l}{\textbf{0.5854}} & \multicolumn{2}{l}{\textbf{0.7809}} \\
\multicolumn{1}{l|}{INDRA} & \multicolumn{1}{l|}{}                       & \multicolumn{3}{c}{0.0143}          & \multicolumn{3}{c}{0.0000}           & \multicolumn{3}{c}{\textbf{0.3766}} & \multicolumn{3}{l}{0.3261}          & \multicolumn{2}{l}{0.6192}          \\ \hline
\multicolumn{1}{l|}{TCN}   & \multicolumn{1}{l|}{\multirow{2}{*}{ID 10}} & \multicolumn{3}{c}{\textbf{0.3706}} & \multicolumn{3}{c}{\textbf{0.1949}} & \multicolumn{3}{c}{0.000}           & \multicolumn{3}{l}{0.0212}          & \multicolumn{2}{l}{0.2036}          \\
\multicolumn{1}{l|}{INDRA} & \multicolumn{1}{l|}{}                       & \multicolumn{3}{c}{0.1779}          & \multicolumn{3}{c}{0.1668}          & \multicolumn{3}{c}{\textbf{0.9413}} & \multicolumn{3}{l}{\textbf{0.0386}} & \multicolumn{2}{l}{\textbf{0.2224}} \\ \hline
\end{tabular}}
\label{table:precision_syncan}
\end{table*}
\vfill
\noindent\textbf{CrySyS. }The message IDs in the SynCAN dataset contains signals that are physically interdependent, but are very weakly correlated; this also increases the difficulty of the detection task. In order to assess how the two models perform in a different setting, we consider two message IDs of the CrySyS dataset which contains more signals with a strong correlation. Here, similarly to SynCAN, only one signal was attacked. The results are shown in Table~\ref{table:crysys_results}. We notice that both models still achieve high accuracy and a low false positive rate, with TCN showing a high precision for both attacks, as opposed to INDRA, failing to detect the attack in message $280$. \begin{table}[tb]
\caption{Results for the CrySyS dataset.}
\centering
\scalebox{0.9}{
\begin{tabular}{l|l|ccc}
\hline
Model & Data                    & Acc.       & FPR             & Precision                           \\ \hline
TCN   & \multirow{2}{*}{ID 280} & \textbf{0.8833} & 0.0426          & \multicolumn{1}{c}{\textbf{0.7766}} \\
INDRA &                         & 0.7989          & \textbf{0.0000} & \multicolumn{1}{c}{0.0000}          \\ \hline
TCN   & \multirow{2}{*}{ID 290} & \textbf{0.9159} & 0.0687 & 0.7701                              \\
INDRA &                         & 0.8617          & \textbf{0.0378}          & \textbf{0.7755}                     \\ \hline
\end{tabular}}
\label{table:crysys_results}
\end{table}

\begin{figure}[tb]
\centering
\includegraphics[width=0.95\linewidth]{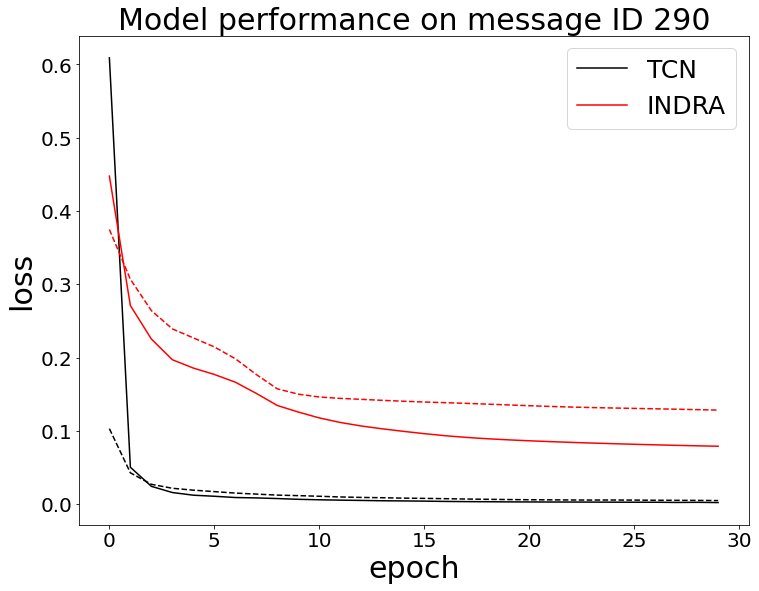}
\caption{Training loss (continuous lines) and validation loss (dashed lines) of the two models on message ID 290 of the CrySyS dataset.}
\label{fig:loss_290}
\end{figure}

\noindent\textbf{Summary. }The simple TCN architecture achieves a slightly better accuracy compared to the INDRA model on both datasets. 
A remarkable achievement of TCN is the significant reduction of false positives (by a factor of 10) in nearly all cases: this translates to a more reliable detector in practice. Another advantage of the TCN is that it is quick to train, and achieves in general lower training and validation loss (see Figure \ref{fig:loss_290} for an example).  

\section{\uppercase{Conclusions}}
\label{sec:conclusion}
In this paper
we examined the applicability of temporal convolutional networks to CAN bus intrusion detection, with a focus on message modification attacks. To this end, we proposed a lightweight TCN, and showed that its classification performance compared favorably to the state-of-the-art baseline INDRA across different datasets and attack classes.
Specifically, we demonstrated that our computation-efficient and compact TCN model achieves similar or better accuracy, while reducing false positives with an order of magnitude. This shows that TCNs have a great potential both in modeling CAN bus signal and being deployed in practical settings.

\noindent\textbf{Future work. }
First of all, the elimination of the early termination mechanism would potentially yield better performance; early termination was necessary in our experiments due to hardware-related constraints.
Second, the TCN architecture was kept very simple on purpose to ensure a computationally lightweight model. However, the learning abilities of the network could be improved by increasing the filter size and the dilation factor between causal convolutions, and by stacking additional residual blocks together. Third, it is worthwhile to investigate how message-based and signal-based intrusion thresholds, and the underlying intra-message signal correlation influence the performance of both models for different attack classes. 
Finally, correlations between signals across different message IDs could be considered leading to a more accurate representation of normal CAN bus behaviour. To this end, an architecture combining multiple TCN blocks (modeling individual message IDs a la CANet~\cite{Hanselman2020}) could be used. 

\section*{\uppercase{Acknowledgements}}
This work has been partially funded by the European Commission via the H2020-ECSEL-2017 project SECREDAS (Grant Agreement no. 783119).
The research presented in this paper and carried out at the Budapest University of Technology and Economics have been supported by the NRDI Office, Ministry of Innovation and Technology, Hungary, within the framework of the Artificial Intelligence National Laboratory Programme, and the NRDI Fund based on the charter of bolster issued by the NRDI Office.

\bibliographystyle{apalike}
{\small
\bibliography{example}}

\end{document}